\documentclass[preprint]{aastex}
\usepackage{graphicx}
\usepackage{lscape}

\usepackage{times}
\usepackage{epsfig}
\usepackage{epsf}
\usepackage{rotating}
\begin{document}


\title{High resolution NaI and CaII absorption observations towards M13, M15 and M33.}


\author{
Barry Y. Welsh\altaffilmark{1},
Jonathan Wheatley \altaffilmark{1}
and Rosine Lallement \altaffilmark{2}}

\altaffiltext{1}{Space Sciences Laboratory, University of California, 7 Gauss Way, Berkeley, CA 94720} \altaffiltext{2}{Service d'Aeronomie du CNRS, 91371, Verrieres-le-Buisson, France}



\begin{abstract}
We present high resolution (R = 60,000) measurements of the NaI D1 $\&$ D2 (5890\AA)
and CaII K (3933\AA)
interstellar absorption line profiles recorded towards several post-AGB stars located within the M13 and M15 globular clusters, supplemented with a lower resolution spectrum of the CaII K-line observed in absorption towards an Ofpe/WN9 star in the central region of the M33 galaxy.
The normalized interstellar absorption
profiles have been fit with cloud component velocities, doppler widths and column densities in order to investigate the kinematics and physical conditions of the neutral and partially ionized gas observed along each sight-line.
Our CaII observations towards M13 have revealed 4 absorption components that can be identified with galactic Intermediate Velocity Clouds (IVCs) spanning the -50 $>$ V$_{lsr}$ $>$ -80 km s$^{-1}$ range. The NaI/CaII ratio for these IVC's is $<$ 0.3, which characterizes the gas as being warm (T $\sim$ 10$^{3}$K) and partially ionized.
Similar observations towards two stars within M15 have revealed absorption due to a galactic IVC at V$_{lsr}$ $\sim$ +65 km s$^{-1}$. This IVC is revealed to have considerable velocity structure, requiring at least 3 cloud components to fit the observed NaI and CaII profiles. 

CaII K-line observations of a sight-line towards the center of the M33 galaxy have revealed at least 10 cloud components. A cloud at V$_{lsr}$ $\sim$ -130 km s$^{-1}$ is either an IVC associated with the M33 galaxy occurring at +45 km s$^{-1}$ with respect to the M33 local standard of rest, or it is a newly discovered HVC associated with our own Galaxy. In addition, 4 clouds have been discovered in the -165 $>$ V$_{lsr}$ $>$ -205 km s$^{-1}$ range. Three of these clouds are identified with the disk gas of M33,  whereas a component at -203 km s$^{-1}$ could be IVC gas in the surrounding halo of M33. 
\end{abstract}



\maketitle


\section{Introduction}
It has long been known that the Galaxy is surrounded by ionized and neutral gas
clouds in the form of a hot and ionized galactic halo  \citep{spitz56}, in addition to the
numerous neutral intermediate-velocity (IV) and high-velocity (HV)
gas clouds \citep{wak97}. These two types of cloud
are normally defined by their observed radial
velocities, such that IVCs have velocities of $\mid$ V$_{LSR}$ $\mid$ 30 - 90 km s$^{-1}$
and HVCs have velocities of $\mid$ V$_{LSR}$ $\mid$ $>$ 90 km s$^{-1}$.
Understanding
the physical and chemical characteristics of both types of gas cloud is fundamental in defining
our current ideas on how matter is exchanged between the Galaxy and the
surrounding intergalactic medium. Although the origins of both IVCs and HVCs
are still hotly debated  \citep{spit08}, it seems likely that they are linked
to either expelled gas from a supernova-driven galactic fountain \citep{breg80} or
they represent accumulations of tidally-stripped or condensed and 
cooled gas \citep{put03, mall04}.
Irrespective of their
true origin, it is generally agreed that IVCs and HVCs are
providing infalling gas, which is theorized to be
required in order to sustain both the star formation process
and the presence of an interstellar medium in our Galaxy \citep{van62}.
If IVCs and HVCs are a major source of  this gaseous infall, then knowledge
of the mass and chemical composition of each cloud are key input parameters for any model
of galactic evolution and star formation  \cite{alib01}. 

The chemical composition and physical state of
gas in HVCs and IVCs are best determined through absorption
measurements at ultraviolet wavelengths  \citep{wak01,richt01,coll07}. These data
suggest that the majority of  neutral HVCs have subsolar metallicities, implying that some fraction
of the gas is of a primordial extragalactic origin. However recent observations of a
new type of highly ionized
HVCs that are not
associated with neutral HI emission, have revealed supersolar
metallicity values which suggests a galactic origin for the gas in
these particular HVCs \citep{fox06, zech08}.
Similarly,
the gas associated with IVCs also appears to
possess near solar abundances \citep{richt01}, except that the majority
of IVCs are relatively nearby objects of distance $<$ 2kpc compared with the
more distant HVCs of
distance $>$ 5 kpc  \citep{olan08}. However, we note that all these assumptions are
based on relatively few observations of HVCs and IVCs, with the data often being of a low
S/N ratio because
of the relative faintness of the background sources used in the absorption observations.

However, with the advent of high resolution spectrographs coupled to very large
aperture ground-based telescopes,
it is now possible to routinely investigate the interstellar gas residing beyond our own Galaxy through
classical absorption spectroscopy techniques that involve distant (d $>$ 5 kpc) background stellar sources. In particular, the well-known interstellar lines of NaI-D (5890\AA)
and CaII-K (3933\AA) have been extremely useful in revealing
the presence of, and determining
accurate distances to, IVCs and HVCs \citep{smok04, smok07, kenn98}. In particular, interstellar
absorption observations of the sight-lines towards hot post-AGB stars within globular
clusters of known distance has proven highly fruitful.  Their sight-lines
can provide (a) distance limits
to absorption arising within foreground IVCs and HVCs as seen
in their visible NaI and CaII
interstellar absorption spectra  \citep{lehn99} and (b) metal abundance determinations of foreground
IVC and HVC gas, through interstellar absorption spectra of numerous
ultraviolet resonance lines \citep{zech08}.

In this Paper we present high resolution (R $\sim$ 60,000) observations of the
NaI D-lines and the CaII K-line seen in absorption towards two post-AGB stars
in the globular cluster NGC 7078 (M15) and toward one post-AGB star in NGC 6205 (M13).
We also present a lower resolution CaII K-line spectrum of the hot star UIT-236
which resides in a planetary nebula in the central regions of the M33 galaxy.
Using these data we discuss the velocity structure of the interstellar absorption observed
along each sight-line and attempt to associate these cloud components with a
galactic,  extended halo or extra-galactic origin.
These data will also supplement a forthcoming abundance analysis of these 4 sight-lines using
ultraviolet absorption observations (with a resolving power of $\sim$ 20,000) using the Cosmic Origins Spectrograph
due to be flown in mid-2009 on Servicing Mission 5 to the Hubble Space Telescope.

\section{Observations and Data Reduction}
We present absorption observations of the interstellar NaI D-line doublet at $\sim$ 5890\AA\ and
the CaII K-line at 3393\AA\  recorded towards hot post-AGB stars in the M13 (NGC 6205) and 
M15 (NGC 7078) globular clusters,
in addition to the Of/WN9 star UIT-236 in the M33 galaxy. The 4 target stars, together with
associated astronomical information and appropriate references, are listed in Table 1. 
These data were obtained during service observing runs performed during 2008
March to July with the
High Dispersion Spectrograph \citep{nog02} on the 8.2m Subaru telescope of
the National Astronomical Observatory of Japan in Hawaii. The spectrograph was configured
with a cross-disperser angle of 5.52$^{\circ}$ to receive light passing through a 0.6 arc sec
wide x 5 arc sec long slit, with the spectra being recorded on two 4k x 2k pixel EEV CCD's.
Spectra were recorded separately for the NaI lines (using the StdYb instrumental set-up) and
the CaII line (using the StdBc instrumental set-up). Unfortunately this set-up configuration
precluded measurement of the CaII H-line at 3368\AA, which appeared in
an echelle order very close to the edge
of the detector.

The raw data were reduced in a similar manner to that described in Sfeir et al. \cite{sfeir99},
which includes cosmic ray removal, CCD bias subtraction, flat-fielding and inter-order background
subtraction. The spectra were wavelength calibrated using Th-Ar calibration lamp spectra, which
resulted in a wavelength accuracy of $\sim$ 2 km s$^{-1}$. The spectral resolution
of the M13 and M15 spectra was determined to be 5 km s$^{-1}$, whereas the resolution
of the spectrum recorded towards M33 (which was recorded with 2 pixel binning) was 9 km s$^{-1}$. 
The telluric water vapor lines that contaminate the NaI absorption spectra
were removed using a synthetic transmission spectrum described
in Lallement et al. \cite{lall93}. The spectra were all well-exposed
with typical S/N ratios in excess of 20:1. Finally, the spectral data were converted into
velocities in the local standard of rest (LSR) frame.

The calibrated spectra were fit with a high order polynomial
to produce local continua in order to establish normalized residual intensities for
the absorption lines of interest.
These normalized profiles were then fit with multiple absorption components
(i.e. interstellar clouds) using a line-fitting program described in Sfeir et al. \cite{sfeir99}.
This program assigns a 3-parameter model fit to the observed profiles using values
for the interstellar gas cloud velocity, V, a Gaussian velocity dispersion, $\it b$ and a cloud
component column density, $\it N$. For the case
of the NaI D1 and D2 line-doublet, the best-fit was performed
simultaneously on both line profiles. Fits were carried out using the $\it minimum$ number
of absorption components, with the addition of extra components only being deemed necessary
if the $\chi$$^{2}$ residual error between the observed and model data points decreased
by more than 11.1  \citep{vall93}. In order to derive meaningful
interstellar gas temperatures for these fits, the $\it b$-values were constrained to
be $<$ 2.9 km s$^{-1}$
 ($<$20,000K ) for the CaII components and to be $<$3.3 km s$^{-1}$
($<$ 15,000K) for the NaI components.
The resultant best-fit values of V$_{lsr}$, $\it b$ and
$\it N$ for the NaI and CaII absorption lines are listed in Table 2, and in Figure 1 we show the
observed spectra and their respective model fits.
We have also included upper limit estimates for certain important IV
components where no
absorption feature was detected with significance. These values
were derived from conservative estimates of
the strength of a potential absorption feature appearing at a level $>$ 2.5-$\sigma$
above the rms value of the local continuum. The corresponding upper limit on the column density
of such features was determined under the assumption of a linear curve of growth and is
reported in Table 2.

\section{Discussion}
\subsection{The sight-line to Barnard 29 in the M13 globular cluster}
The 7.2kpc sight-line to Barnard 29  in the M13 globular
cluster ($\it l$ = 59$^{\circ}$, $\it b$ = +41$^{\circ}$) passes through the
galactic  Hercules shell (GS 57+41) at a distance
of $\sim$ 150pc  \citep{lil92}, as well as an Intermediate Velocity Cloud (IVC)
first observed in 21cm HI  emission  \citep{kerr72}.
The $\it I.U.E$ observations
of Barnard 29 by de Boer and Savage \cite{deBoer83} revealed IV gas at a 
velocity centered at $\sim$ 80 km s$^{-1}$ in the MgII, CII and OI absorption lines.
Due to the similar velocities of the UV and the HI data it was argued that this absorption
arises at an unknown distance within the galactic halo.
More recent
Leiden-Dwingeloo HI observations  have revealed the
IV cloud to possess a central velocity of -64.8 km s$^{-1}$ \citep{smok06}, 
and the HI maps
of  Kuntz $\&$ Danly \cite{kuntz96} show that M13 lies just off the boundary that
defines the spatial extent of the IV Arch, a feature that extends over a large portion of the 
northern galactic hemisphere. 
Previous NaI D-lineand CaII K-line observations of sight-lines
towards several stars within M13 (but not Barnard 29) 
have revealed IV absorption over the -65 to -72 km s$^{-1}$ range \citep{shaw96}. 
Observations of the CaII K-line in the sight-line towards Barnard 29 by Bates et al. \cite{bates95}
have shown a blend of at least three IV absorption
components spanning the -40 to -90 km s$^{-1}$ range, whereas
the more recent measurements of Smoker et al. \cite{smok06} have revealed a similarly broad absorption feature
that was fit with only one component centered at V$_{lsr}$ = -48.4 km s$^{-1}$.
Both sets of data are consistent with
our present CaII-K observations that reveal 4 components at V$_{lsr}$ = -50.1, -59.6,
-70.1 and -78.6 km s$^{-1}$.
Our NaI D-line spectra show no measurable absorption at velocities in the -40 to -90 km s$^{-1}$
range (to a detection level of $<$ 5 m\AA\ for the D2 line), in contrast
with other (lower S/N) absorption observations of 
sight-lines towards 3 angular close stars in M13 \citep{shaw96}.

Our integrated value of CaII column density for all 4 IV components
is log N(CaII) = 11.64 cm$^{-2}$,
which can be compared with the similar values of log N(CaII) = 11.65 and 11.92 cm$^{-2}$ 
measured respectively by Smoker et al \cite{smok06} and Bates et al. \cite{bates95}. 
Additionally, we note in Figure 3 of Smoker et al. \cite{smok06} there appears to be absorption
features in both of the D2 and D1 line spectra of Barnard 29 in the -20 to -80 km s$^{-1}$ range.
Unfortunately the velocities of these features are not common to both NaI lines, suggesting that they
may be due to incomplete removal of telluric water vapor lines in this wavelength region. If this is 
the case, then both of our NaI D-line profiles argue against any measurable IV absorption
formed over the same velocity range as the CaII-K line.
There is a large  ($\sim$ factor of 100) variation in the abundance of Na for high- and intermediate-velocity
gas compared with that measured for low-velocity gas \citep{wak01}, which may explain why, although NaI
IV components have been detected along several other
sight-lines to M13 \citep{shaw96}, we have not been able to detect similar IV NaI towards Barnard 29.

The NaI/CaII ratio for all four of the presently detected IV components is $<$ 0.3, which is a value found widely
for interstellar clouds that can be characterized as being
warm (T $\sim$ 10$^{3}$K), of low density (n$_{H}$ $<$ 1 cm$^{-3}$ and partially
ionized \citep{hobbs75}. It has also been shown that  the NaI/CaII ratio
does not depend on velocity for sight-lines through IVCs and HVCs, since the
ratio depends mostly on N(HI) which has lower values at higher velocities \citep{wak01}.
The physical conditions of this IV gas can be compared with the
main absorbing cloud at V$_{lsr}$ $\sim$ +10 km s$^{-1}$, whose higher NaI/CaII ratio
of 3.0 is typical for clouds that are present in colder, more neutral and denser interstellar regions.

As discussed in the Introduction measurement of the distance to IVCs and HVCs
is an important parameter. We are able to place a lower limit to the distance of the IV
absorption of d = 265pc, based on CaII absorption measurements towards the star
HD 156633 ($\it l$ = 56.4$^{\circ}$, $\it b$ = +33.1$^{\circ}$) which does not reveal any measurable
absorption components for V$_{lsr}$ $<$ - 30 km s$^{-1}$ \citep{welsh09}).
The upper limit of d = 7.2kpc is obtained from the distance to the M13 globular cluster.
Although we did not detect any IV absorption in the NaI lines, we note that this is also consistent
with the NaI measurements towards the foreground star HD 147113 ($\it l$ = 61$^{\circ}$, $\it b$ = +46$^{\circ}$)
of distance $\sim$600pc \citep{lil92}.

\subsection{The sight-lines to K648 and Zng1 in the M15 globular cluster}
Both of the 10.3 kpc long sight-lines towards the M15
globular cluster ($\it l$= 65$^{\circ}$, $\it b$=-27.3$^{\circ}$)
pass through a previously known IVC with V$_{lsr}$ $\sim$ +70 km s$^{-1}$ \citep{cohen79,
kerr72}. More recent, higher resolution optical and HI observations of this
IV gas (often called the `g1' cloud) have resulted in a distance estimation
of between 1.8 to 3.8 kpc to this IVC
\citep{wakker08, smoker02a, smoker02b}. The
absorbing gas within this cloud complex
has been shown to exhibit significant small-scale structure with NaI column densities
varying by up to a factor 16 across the foreground cloud \citep{meyer99}.

The star Zng-1 has been previously observed at a spectral resolution of $\sim$ 7.5 km $^{-1}$
in both the NaI and CaII absorption lines \citep{smoker02a, smoker02b}. These data
revealed appreciable low velocity absorption covering the -30 to +20 km s$^{-1}$ range,
which can be associated with gas in the galactic disk (and perhaps lower halo). 
An absorption feature centered
at V$_{lsr}$ = +64 km s$^{-1}$ was observed in both the NaI and CaII lines, which matched
the velocity of the HI emission from the foreground IVC gas recorded towards M15. An extra absorption component at V$_{lsr}$ = +53 km s$^{-1}$
was also tentatively suggested by these visible observations. Under the assumption that a value of 
the HI column density to Zng-1 is N(HI) = 5 x 10$^{19}$ cm$^{-2}$, the authors derived 
values of N(CaII) / N(HI) $\sim$ 5 x 10$^{-8}$ and
N(NaI) / N(HI) $\sim$ 1.3 x 10$^{-8}$ for the entire IV gas cloud.

The star K648, which is thought to be the central star of a planetary nebula lying within
M15, has been observed in the ultraviolet and has been
shown to possess a significant stellar wind with the nebula expanding at $\sim$ 15 km s$^{-1}$
with respect to the ambient medium \citep{bianch01} . No previous visible interstellar absorption observations
exist for this star.

Our present higher resolution NaI and CaII  measurements towards both Zng-1 and K648
(which are separated by only 1.25 arc minute on the sky)
show very similar patterns of visible absorption (see Figure 1). These profiles
are dominated by low velocity ( -30 to +20 km s$^{-1}$)
absorption originating in the galactic disk/halo, in addition to an IVC absorption
centered at V$_{lsr}$ $\sim$ +65 km s$^{-1}$. The low-velocity galactic components
generally have NaI/CaII ratios $>$ 1.0, consistent with the sight-lines
sampling a cold and neutral diffuse
interstellar medium.
In contrast, 
the NaI/CaII ratios for all of the components associated with the IVC have values $<$ 1.0,
suggestive of a warm, partially ionized and low density interstellar cloud medium.
Our data also clearly show that the IVC gas 
sampled towards both stars has considerable velocity structure, 
requiring at least 3 cloud components to fit the observed NaI and CaII absorption profiles.
This structure of the IVC, in which several clouds with similar velocities
are present (i.e. +55 $<$ V$_{lsr}$ $<$ +75 km s$^{1}$), is consistent with the theoretical
scenario in which possible cloud-to-cloud collisions could result in triggering the star
formation process within such complexes \citep{christ97}.

\subsection{The sight-line towards the star UIT-236 in the M33 galaxy}
M33, the Triangulum  spiral galaxy (NGC 598), is a member of the Local Group
lying at a distance of $\sim$ 820 kpc in the direction ($\it l$= 133.6$^{\circ}$, $\it b$=-31.3$^{\circ}$)
and is about one quarter the size of both the Milky Way and
Andromeda galaxies. It has a systemic velocity of V$_{lsr}$ $\sim$ -180 km s$^{-1}$ and is viewed
near face on. The metallicity of the galaxy is sub-solar, with a radial abundance gradient that differs
for different elements \citep{rosol08}. 

The interstellar sight-line towards 
the Of-type star  UIT-236, which is part of the NGC 588 giant HII region, is close to the nucleus of the galaxy and low
resolution ultraviolet observations have 
revealed significant mass-loss (through P-Cygni wind profiles) from this star \citep{bianch04}.  Low
resolution far ultra-violet absorption
observations of UIT-236 recorded with $\it FUSE$ have revealed the
presence of both hot 300,000$^{\circ}$K OVI  and
lower temperature CII gas at V$_{lsr}$ $\sim$ 180 km s$^{-1}$  that 
can be definitely associated with the host galaxy's
ISM \citep{wak03,hutch04}. Observations of interstellar molecular H$_{2}$ and several
low ionization ions towards the core of NGC 588 (i.e.
UIT-236) with $\it FUSE$ have revealed a common absorption at V$_{lsr}$ $\sim$ -140 km s$^{-1}$
 \citep{bluhm03}. This is a very similar absorption velocity to that of the `weak'  
component seen in 21cm HI emission towards M33 by Rogstad et al. \cite{rog76}.

Our present CaII-K absorption observations towards UIT-236 shown
in Figure 1 reveal five major
absorption systems centered at V$_{lsr}$ $\sim$ -205, -175, -130, -35 and 0 km s$^{-1}$.
Unfortunately we have no accompanying interstellar NaI observations towards this star.
However, we can associate the two sets of components observed at V$_{lsr}$ = 0 and
 -35 km s$^{-1}$ with
absorption due to foreground galactic (and possibly inner galactic halo) interstellar gas. We note
that the component at V$_{lsr}$ $\sim$ -35 km s$^{-1}$ is consistent with the velocity of gas
associated with the outer region of the IVC HI gas which has been
termed the `PP Arch' \citep{wak01}. This stream of IV gas has an implied distance of
1.0 to 2.7 kpc with velocities in the -30 to -60 km s$^{-1}$ range  \citep{wakker08}.
If we assume an integrated HI column density of N(HI) $\sim$ 5 x 10$^{19}$ cm$^{-2}$ for the IV gas
(-25 to -45 km s$^{-1}$) in this direction, then we obtain a N(CaII)/N(HI) ratio of $\sim$ 6 x 10$^{-8}$
for the IV cloud complex. Wakker $\&$ Mathis \cite{wakk00} discovered an empirical
relationship between observed values of N(CaII) and N(HI) along
many sight-lines, such that for the PP Arch IVC they predict a
 N(CaII)/N(HI) ratio that is slightly larger than our observed value. This supports
 the current notion that IVC's seem deficient in ionized calcium compared
 with HVCs. 
 
 The cloud component observed at V$_{lsr}$ $\sim$ -130 km s$^{-1}$ poses the question as to whether
 it is a high negative velocity HVC belonging to the Milky Way galaxy, or whether it is an IVC associated
 with the M33 galaxy that occurs at $\sim$ +45 km s$^{-1}$ with respect to the M33 local
 standard of rest? The observed velocity of this absorption component is very similar
 to that of both UV molecular H$_{2}$ and several
low ionization lines measured towards NGC 588 with
$\it FUSE$ \citep{bluhm03}. This is also very similar velocity to that of the `weak'  
component seen in  21cm HI emission by Rogstad et al. \cite{rog76}, who theorized that
this absorption could be due to gas residing above and below the plane of the M33 galaxy.
In addition the all-sky map of galactic HVC's
shows only highly negative (V $<$ -300 km s$^{-1}$) velocity gas present in the general
direction of M33 \citep{wakker08}. We have not been able to confirm
the presence of such HV gas with our present interstellar 
observations to a level of N(CaII) $<$ 10$^{11}$ cm$^{-2}$. Until future, more extensive, UV
absorption observations of the
V$_{lsr}$ $\sim$ -130 km s$^{-1}$ component are performed, we favor this component's
association with an IVC of the M33 galaxy.

The interstellar CaII gas clouds observed
in absorption with velocities in the range  -175 $>$ V$_{lsr}$ $>$ - 200 km s$^{-1}$ can be directly associated with gas
of the M33 galaxy, since such velocities are close to the systemic velocity of the host galaxy.
As noted previously, the FUV absorption lines of 
interstellar molecular hydrogen, CII (1036\AA), FeII (1144\AA), ArI (1048\AA) and OVI (1032\AA) have also been detected
with similar velocities towards M33
\citep{bluhm03,hutch04}. 
The cloud component at V$_{lsr}$ $\sim$ -173 km s$^{-1}$ is saturated and is therefore most probably
associated with gas in the disk of M33, which we are viewing almost face-on. The HI emission
maps of M33 by Rogstad et al. \citep{rog76} show the presence of the main velocity
components   (-150 to - 200 km s$^{-1}$) occurring in the region within $\pm$5 arc min of the
galaxy center. 

The component with V$_{lsr}$ = -203 km s$^{-1}$ could be an IVC of M33 with a relative
velocity of $\sim$ -30 km s$^{-1}$ wrt the disk gas.  We note that Grossi et al. \cite{gross08}
have recently found several HI clouds surrounding M33 that fall into this velocity range. The current
view of M33 is that it is a satellite of the larger M31 galaxy and the
complex of surrounding HI (and HII) clouds
may be either debris flowing into M33 from the IGM or from a previous interaction with M31.
Gaining gas phase abundances of this component (from forthcoming UV
absorption observations)  would be extremely informative with
respect to a better determination of the origin of this cloud.

\section{Conclusion}
We have presented high resolution (R = 60,000) absorption measurements of the interstellar lines of  NaI D1 $\&$ D2 (5890\AA)
and CaII-K (3933\AA) recorded towards post AGB stars in the M13 and M15 globular clusters, supplemented with a lower resolution spectrum of the CaII-K line seen towards an Ofpe/WN9
star in the central region of the M33 galaxy. The absorption line-profiles have been fit with
cloud component velocities and column densities such that the kinematics
and physical conditions of the neutral and partially ionized gas components can be
investigated.

Four CaII-K absorption components that can be identified with galactic IVC's spanning the -50 to -80 km s$^{-1}$ range have been detected towards the M13 globular cluster. The associated NaI/CaII column
density ratio for this IVC gas is $<$ 0.3, which suggests that the gas is warm (T $\sim$ 10$^{3}$K) and partially ionized. The observations of two stars within M15 have revealed absorption due to
a galactic IVC at V$_{lsr}$ $\sim$+65 km s$^{-1}$, which requires at least 3 separate cloud components
(closely spaced in velocity) to adequately fit the NaI and CaII profiles.

CaII K-line observations of the sight-line to the M33 galaxy have revealed at least 10 gas cloud
components. A cloud at V$_{lsr}$ $\sim$ -130 km s$^{-1}$ is either an IVC associated
with the M33 galaxy (occurring at +45 km s$^{-1}$ with respect to the M33 local standard of rest), or
it is an newly discovered HVC associated with our own Galaxy. Also, 4 gas clouds have
been detected in the -165 to -205 km s$^{-1}$ range, of which three are probably
associated with the disk gas of M33. The cloud at V$_{lsr}$ = -203 km s$^{-1}$ could be an
IVC residing in the surrounding halo of M33.

Finally, we note that these high resolution visible data will be extremely useful in providing
sight-line velocity templates for the forthcoming lower resolution UV absorption studies of
these same stars to be carried out with the newly installed  Cosmic Origins Spectrograph on
the Hubble Space Telescope in mid-2009.

\begin{acknowledgements}
We particularly acknowledge the dedicated
team of engineers, technicians, and research staff who recorded these data with
the Subaru Telescope, which is operated by the National Astronomical
Observatory of Japan.
This publication makes use of data products from the SIMBAD database,
operated at CDS, Strasbourg, France.  BYW acknowledges funding for this
research through the NSF award AST-0507244.
\end{acknowledgements}

\begin{table*}
\caption{Stellar Target Information}
\label{table:1}
\centering
\begin{tabular}{c l l c c c c c c}
\hline
\hline
Star & R.A. (2000.) & Decl(2000.) & m$_{v}$ & T$_{eff}$ (K) & E(B-V)& distance (kpc) & Reference \\
\hline
NGC 6205: Barnard 29&16:41:+34.0&+36:26:08.2&13.14& 20,000&0.02& 7.2& (1) \\
NGC 7078: K648&21:29:59.4&+12:10:26.0&14.7&39,000&0.13&10.3& (2) \\
NGC 7078: Zng1&21:29:58.1&+12:11:44.2&14.8&28,000&0.13&10.3& (3) \\
M33: UIT 236&01:33:53.6&+30:38:51.8&18.1&33,000&0.10&820& (4) \\
\hline
\hline
\multicolumn{8}{l}{\ (1) = \citep{dix98}, \ (2) = \citep{bianch01}, \ (3) = \citep{moon04}, \ (4) \citep{bianch04}}\\
\hline
\hline
\end{tabular}
\end{table*}
\begin{table*}[htbp]
\caption{NaI and CaII Absorption Line Best-Fit Parameters (stars listed by increasing distance)}
\label{table:2}
\centering
\scriptsize
\begin{tabular}{ccccccccccc}
\hline
\hline
Star&V$_{lsr}$&$\it b$&N&&&V$_{lsr}$&$\it b$&N& NaI/CaII \\
&km s$^{-1}$&&(10$^{10}$ cm$^{-2}$)&&&km s$^{-1}$&&(10$^{10}$ cm$^{-2}$)& \\
\hline
{\bf \object{M13--Barnard 29}} &&&& \\
...NaI...&&&$<$2.5&&...CaII...&-78.6&2.9&12.2$\pm1$&$<$0.20 \\
&&&$<$ 2.5&&&-70.0&2.8&12.4$\pm1$&$<$0.20 \\
&&&$<$2.5&&&-59.5&2.8&8.4$\pm1$&$<$0.3 \\
&&&$<$2.5&&&-50.0&2.9&10.5$\pm1$&$<$0.24\\
&&&&&&-11.5&2.9&26.7$\pm3$&- \\
&-6.5&3.3&8.5$\pm1$&&&-6.6&2.9&43.0$\pm5$&0.20 \\
&+2.2&3.3&11.5$\pm1.5$&&&+0.5&2.9&66.0$\pm8$&0.17 \\
&&&&&&+5.0&2.9&52.0$\pm6$&- \\
&+11.4&3.3&121.5$\pm15$&&&+9.3&2.9&40.2$\pm5$&3.0 \\
{\bf \object{M15--K648}} &&&& \\
...NaI...&&&$<$5&&...CaII...&-30.4&2.9&23.3$\pm1$&$<$0.21 \\
&&&$<$ 4&&&-22.3&2.9&92.7$\pm5$&$<$0.04\\
&-17.1&3.3&15.3$\pm3$&&&-15.2&2.9&142$\pm20$&0.11\\
&-8.5&0.3&84.4$\pm10$&&&-6.0&2.9&188$\pm20$&0.45\\
&-2.0&3.3&560**&&&-0.5&2.9&147$\pm18$&$>$ 3.8\\
&+9.5&3.0&555**&&&+7.2&2.9&153$\pm18$&$>$3.6\\
&&&&&&+17.0&2.9&34.9$\pm4$&-\\
&&&$<$4&&&+53.6&2.9&22.3$\pm5$&$<$0.18 \\
&+60.8&3.3&17.2$\pm3$&&&+60.1&2.9&38.8$\pm4$&0.44 \\
&+68.8&3.2&32.3$\pm5$&&&+68.4&2.9&238$\pm15$&0.14 \\
{\bf \object{M15--Zng1}} &&&& \\
...NaI...&&&$<$3&&...CaII...&-31.5&2.9&25.8$\pm3$&$<$0.12 \\
&&&$<$3&&&-23.7&2.8&65.9$\pm8$&$<$0.05 \\
&-16.6&3.0&21.5$\pm3$&&&-17.5&2.9&150$\pm20$&0.14 \\
&-9.5&0.3&21.4$\pm4$&&&-11.0&2.9&53.5$\pm6$&0.4\\
&-1.5&1.1&855**&&&-3.7&2.9&157$\pm12$&$>$5.5\\
&+1.8&3.2&83.6$\pm10$&&&+1.4&2.9&35.4$\pm5$&2.4\\
&&&-&&&+6.9&2.9&127$\pm15$&-\\
&+13.2&2.6&668**&&&+16.0&2.9&29.3$\pm4$&$>$22.8\\
&&&$<3$&&&+56.0&2.2&29.8$\pm2$&$<$0.1\\
&+60.0&1.2&3.8$\pm1$&&&+63.0&2.5&127$\pm15$&0.03\\
&+67.0&1.8&66.3$\pm8$&&&+68.0&1.6&96.2$\pm7$&0.7\\
&+72.1&2.5&12.2$\pm2$&&&+72.5&2.5&14.3$\pm2$&0.85\\
{\bf \object{M33--UIT-236}} &&&& \\
...NaI...&&&&&...CaII...&-203&2.9&42.2$\pm8$&- \\
&&&&&&-183&2.3&347$\pm50$&-\\
&&&&&&-173&2.9&906**&-\\
&&&&&&-165&1.8&36.3$\pm8$&-\\
&&&&&&-130&2.9&55.5$\pm6$&-\\
&&&&&&-40.0&2.9&211$\pm30$&- \\
&&&&&&-29.3&2.9&75.1$\pm9$&- \\
&&&&&&-10.0&2.8&49.2$\pm6$&-\\
&&&&&&-0.6&2.6&93.8$\pm15$&-\\
&&&&&&+12.0&0.7&22.6$\pm4$&-\\
\hline
\multicolumn{9}{l}{\sl ** = saturated component} \\
\hline
\hline
\end{tabular}
%\end{flushleft}
\end{table*}

\begin{figure*}
\center
\plotone{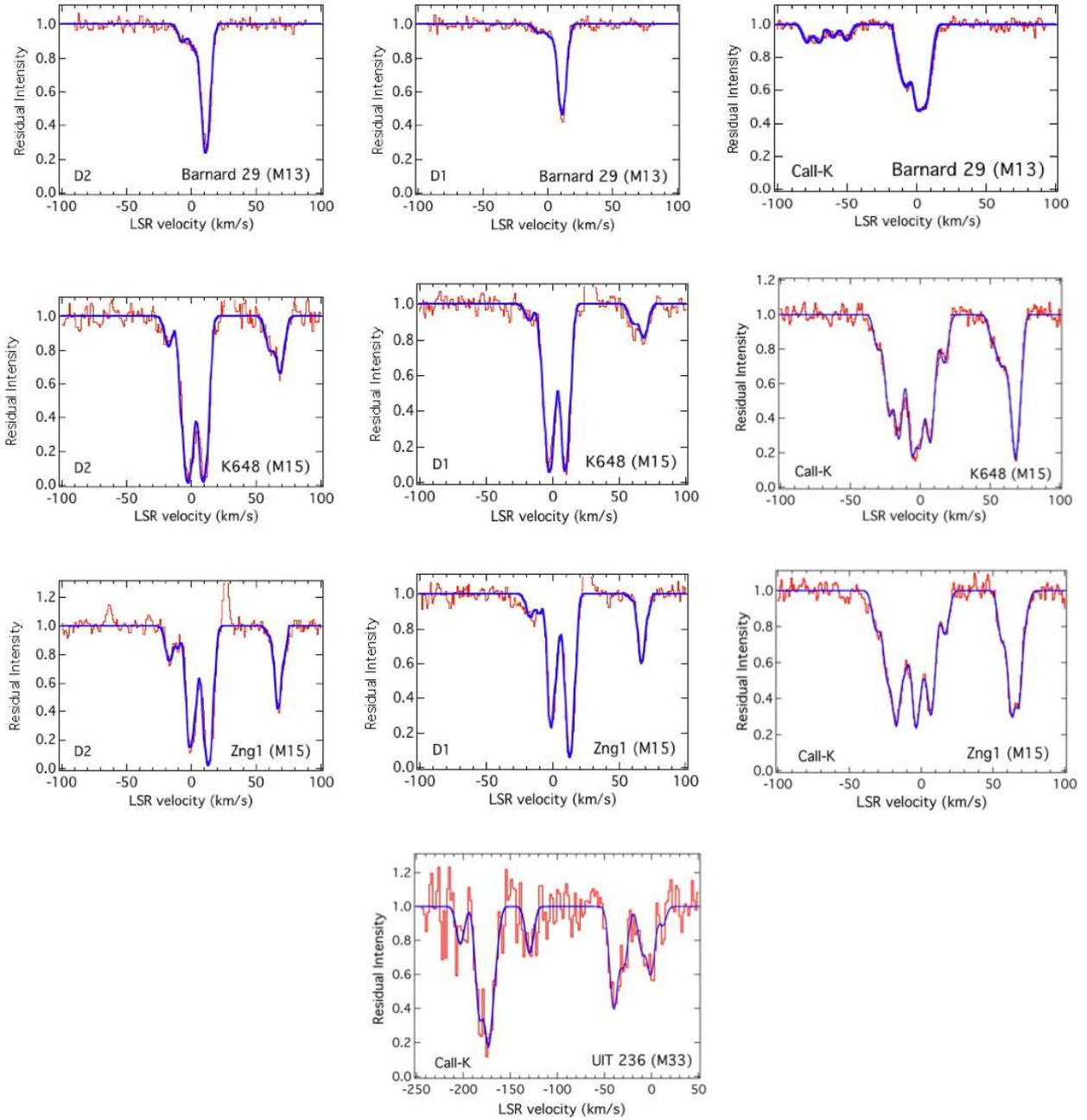}
\caption{Interstellar absorption profiles of the NaI D2, D1 and CaII K-lines observed towards the stars Barnard 29 (M13), K648 (M15), Zng1 (M15) and UIT 236 (M33). Full lines are the model fits superposed upon the normalized residual intensity data points.}
\label{Figure 1}
\end{figure*}


\begin{thebibliography}{}

\bibitem[Alibes et al. 2001]{alib01} Alibes, A., Labay, J. and Canal, R., 2001, \aap, 370, 1103

\bibitem[1995]{bates95} Bates, B., Shaw, C., Kemp, S. et al., 1995, \apj, 444, 672

\bibitem[Bianchi et al. 2001]{bianch01} Bianchi, L., Bohlin, R., Catanzaro, G. et al., 2001, \aj, 122, 1538

\bibitem[Bianchi et al. 2004]{bianch04} Bianchi, L., Bohlin, R. and Massey, P., 2004, \apj, 610, 228

\bibitem[Bluhm et al. 2003]{bluhm03} Bluhm, H., de Boer, K., Marggraf, O. et al., 2003, \aap, 398, 983

\bibitem[Bregman 1980]{breg80} Bregman, N., 1980, \apj, 236, 577

\bibitem[Christodoulou et al. 1997]{christ97} Christodoulou, D., Tohline, J. and Keenan, F., 1997, \apj, 468, 810

\bibitem[Cohen 1979]{cohen79} Cohen, J., 1979, \apj, 231, 751

\bibitem[Collins et al. 2007]{coll07} Collins, J., Shull, M. and Giroux, M., 2007, \apj, 657, 271

\bibitem[1983]{deBoer83} de Boer, K. and Savage, B., 1983, \apj, 265, 210

\bibitem[Dixon $\&$ Hurwitz 1998]{dix98} Dixon, van D. and Hurwitz, M., 1998, \apj, 500, L29

\bibitem[Fox et al. 2006]{fox06} Fox, A., Savage, B. and Wakker, B., 2006, \apjs, 165, 229

\bibitem[2008]{gross08} Grossi, M., Giovanardi, C., Corbelli, E. et al., 2008, \aap, 487, 161

\bibitem[Hobbs 1975]{hobbs75} Hobbs, L., 1975, \apj, 202, 628

\bibitem[Hutchings $\&$ Butler 2004]{hutch04} Hutchings, J. and Butler, K., 2004, \aj, 128, 2234

\bibitem[Kennedy et al. 1998]{kenn98} Kennedy, D., Bates, B. and Kemp, S., 1998, \aap, 336, 315

\bibitem[Kerr $\&$ Knapp 1972]{kerr72} Kerr, F. and Knapp, G., 1972, \aj, 77, 354

\bibitem[1996]{kuntz96} Kuntz, K. and Danly, L., 1996, \apj, 457, 703

\bibitem[1993]{lall93} Lallement, R., Bertin, P, Chassefiere, E. and Scott, N., 1993, \aap, 271, 734

\bibitem[2003]{lall03} Lallement, R., Welsh, B.Y., Vergely, J.L. et al., 2003, \aap\, 411, 447

\bibitem[Lehner et al. 1999]{lehn99} Lehner, N., Rolleston, W., Ryans, R. et al., 1999, \aaps, 134, 257

\bibitem[Lilienthal et al. 1992]{lil92} Lilienthal, D., Hirth, W., Mebold, U. and de Boer, K., 1992, \aap, 255, 323

\bibitem[Maller $\&$ Bullock 2004]{mall04} Maller, A. and Bullock, J., 2004, \mnras, 355, 694

\bibitem[Meyer $\&$ Lauroesch 1999]{meyer99} Meyer, D. and Lauroesch, J., 1999, \apj, 520, L103

\bibitem[Mooney et al. 2004]{moon04} Mooney, C., Rolleston, W., Keenan, F. et al., 2004, \aap, 419, 1123

\bibitem[Noguchi et al. 2002]{nog02} Noguchi, K., Aoki, W., Kawanomoto, S. et al., 2002, PASJ, 54, 855

\bibitem[Olano 2008]{olan08} Olano, C., 2008, \aap, 485, 457

\bibitem[Putman et al. 2003]{put03} Putman, M., Staveley-Smith, L., Freeman, K. et al., 2003, \apj, 586, 170

\bibitem[Richter et al. 2001]{richt01} Richter, P., Sembach, K., Wakker, B. et al., 2001, \apj, 559, 318

\bibitem[1976]{rog76} Rogstad, D., Wright, M. and Lockhart, I., 1976, \apj, 204, 703

\bibitem[Rosolowsky $\&$ Joshua 2008]{rosol08} Rosolowsky, E. and Joshua, S., 2008, \apj, 675, 1213

\bibitem[1999]{sfeir99} Sfeir, D.M., Lallement, R., Crifo, F. and Welsh, B.Y., 1999, \aap, 346, 785

\bibitem[Shaw et al. 1996]{shaw96} Shaw, C., Bates, B., Kemp, S. et al., 1996, \apj, 473, 849

\bibitem[Smoker et al. 2002a]{smoker02a} Smoker, J., Keenan, F., Lehner, N. and Trundle, C., 2002a, \aap, 387, 1057

\bibitem[2002b]{smoker02b} Smoker, J., Haffner, L., Keenan, F. et al., 2002b, \mnras, 337, 385

\bibitem[Smoker et al. 2004]{smok04} Smoker, J., Lynn, B., Rolleston, W. et al., \mnras, 352, 1279

\bibitem[2006]{smok06} Smoker, J., Lynn, B., Christian, D. and Keenan, F., 2006, \mnras, 370, 151

\bibitem[Smoker et al. 2007]{smok07} Smoker, J., Hunter, I., Kalberla, P. et al., 2007, \mnras, 378, 947

\bibitem[Spitoni et al. 2008]{spit08} Spitoni, E., Recchi, S. and Matteucci, F., 2008, \aap, 484, 743

\bibitem[Spitzer 1956]{spitz56} Spitzer, L.J., 1956, \apj, 124, 20

\bibitem[Vallerga et al. 1993]{vall93} Vallerga, J., Vedder, P., Craig, N. and Welsh, B.Y., 1993, \apj, 411, 729

\bibitem[van den Bergh 1962]{van62} van den Bergh, S., 1962, \aj, 67, 486

\bibitem[Wakker 2001]{wak01} Wakker, B., 2001, \apjs, 136, 463

\bibitem[2000]{wakk00} Wakker, B. $\&$ Mathis, J., 2000, \apj, 544, L107

\bibitem[Wakker $\&$ van Woerden 1997]{wak97} Wakker, B.P. $\&$ van Woerden, H., 1997, ARA$\&$A, 35, 217

\bibitem[Wakker et al. 2003]{wak03} Wakker, B., et al., 2003, \apjs, 146,1

\bibitem[Wakker et al. 2008]{wakker08} Wakker, B.P., York, D., Wilhelm, R. et al., 2008, \apj, 672, 298

\bibitem[Welsh $\&$ Lallement 2009]{welsh09} Welsh, B. and Lallement, R., \aap, (in prep)

\bibitem[Zech et al. 2008]{zech08} Zech, W., Lehner, N., Howk, C. et al., 2008, \apj, 679, 460
\end{thebibliography}
\end{document}